\documentclass[journal=cmatex,manuscript=article]{achemso}
\usepackage[usenames,dvipsnames]{color}
\usepackage[english]{babel}
\usepackage{epsfig}
\usepackage{graphicx,subfigure}
\usepackage{float}
\usepackage{tikz}
\usepackage{pgfplots}
\pgfplotsset{compat=1.12}
\usetikzlibrary{calc}
\usetikzlibrary{decorations.markings}
\usetikzlibrary{decorations.pathmorphing}
\usetikzlibrary{positioning,shapes,arrows,decorations.markings,calc}
\usepackage{longtable}
\usepackage{multirow}
\usepackage{arydshln}
\usepackage[colorlinks,citecolor=aaltoBlack,linkcolor=aaltoBlack]{hyperref}
\usepackage{amsmath,bbm,bm,amssymb}
\usepackage[mathcal,mathscr]{eucal}
\DeclareFontFamily{OT1}{pzc}{}
\DeclareFontShape{OT1}{pzc}{m}{it}{<-> s * [1.10] pzcmi7t}{}
\DeclareMathAlphabet{\mathpzc}{OT1}{pzc}{m}{it}
\usepackage[T1]{fontenc}
\usepackage{times}
\usepackage{lmodern}
\usepackage[Euler]{upgreek}
\usepackage{enumitem}
\setenumerate[1]{itemsep=0pt,partopsep=0pt,parsep=0pt,topsep=0pt}
\setitemize[1]{itemsep=0pt,partopsep=0pt,parsep=0pt,topsep=0pt}
\setdescription[1]{itemsep=0pt,partopsep=0pt,parsep=0pt,topsep=0pt}

\usepackage{bookmark}

\usepackage{textcomp}

\usepackage{lscape}

\setcitestyle{numbers,square,citesep={,\kern-.24em}}

\hypersetup{
  colorlinks=false,
  pdfborder={0 0 0},
}

\tikzset{midarrow/.style={decoration={
  markings,
  mark=at position #1 with {\arrow{angle 45}}},postaction={decorate}}
}
\tikzset{waved/.style={decorate,decoration=snake}}

	\definecolor{aaltoBlack}{RGB}{0,0,0}%
	\definecolor{aaltoGray}{RGB}{146,139,129}%
	\definecolor{aaltoRed}{RGB}{237,41,57}%
	\definecolor{aaltoBlue}{RGB}{0,101,189}%
	\definecolor{aaltoYellow}{RGB}{254,203,0}%
	\definecolor{aaltoPurple}{RGB}{102,57,183}%
	\definecolor{aaltoTurquoise}{RGB}{0,168,180}%
	\definecolor{aaltoGreen}{RGB}{0,155,58}%
	\definecolor{aaltoLightGreen}{RGB}{105,190,40}%
	\definecolor{aaltoOrange}{RGB}{255,121,0}%
	\definecolor{aaltoFuchsia}{RGB}{177,5,157}%
\usepackage{soul}
\AddToHook{package/hyperref/after}{\def\htmladdnormallink#1#2{\href{#2}{#1}}}

\SectionNumbersOn

\title{\vspace{-2em} Mixed-halide perovskite alloys \texorpdfstring{$\text{CsPb}(\text{I}_{1-x}^{}\text{Br}_x^{})_3^{}$}{} and \texorpdfstring{$\text{CsPb}(\text{Br}_{1-x}^{}\text{Cl}_x^{})_3^{}\,$}{}: New insight of configuration entropy effect from first principles and phase diagrams}

\author{Fang~Pan}
\affiliation{State Key Laboratory for Manufacturing Systems Engineering; Electronic Materials Research Laboratory, Key Laboratory of the Ministry of Education, School of Electronic Science and Engineering, Xi'an Jiaotong University, Xi'an 710049, China \looseness=-2}

\author{Junni~Zhai}
\affiliation{State Key Laboratory for Manufacturing Systems Engineering; Electronic Materials Research Laboratory, Key Laboratory of the Ministry of Education, School of Electronic Science and Engineering, Xi'an Jiaotong University, Xi'an 710049, China \looseness=-2}

\author{Jinyu~Chen}
\affiliation{State Key Laboratory for Manufacturing Systems Engineering; Electronic Materials Research Laboratory, Key Laboratory of the Ministry of Education, School of Electronic Science and Engineering, Xi'an Jiaotong University, Xi'an 710049, China \looseness=-2}

\author{Lin Yang}
\affiliation{State Key Laboratory for Manufacturing Systems Engineering; Electronic Materials Research Laboratory, Key Laboratory of the Ministry of Education, School of Electronic Science and Engineering, Xi'an Jiaotong University, Xi'an 710049, China \looseness=-2}

\author{Hua~Dong}
\affiliation{Key Laboratory for Physical Electronics and Devices of the Ministry of Education and Shaanxi Key Lab of Information Photonic Technique, School of Electronic Science and Engineering, Xi'an Jiaotong University, Xi'an 710049, China \looseness=-2}

\author{Fang~Yuan}
\affiliation{Key Laboratory for Physical Electronics and Devices of the Ministry of Education and Shaanxi Key Lab of Information Photonic Technique, School of Electronic Science and Engineering, Xi'an Jiaotong University, Xi'an 710049, China \looseness=-2}

\author{Zhuangde~Jiang}
\affiliation{State Key Laboratory for Manufacturing Systems Engineering \& International Joint Laboratory for Micro/Nano Manufacturing and Measurement Technology, Xi'an Jiaotong University, Xi'an 710049, China \looseness=-2}

\author{Wei~Ren}
\affiliation{State Key Laboratory for Manufacturing Systems Engineering; Electronic Materials Research Laboratory, Key Laboratory of the Ministry of Education, School of Electronic Science and Engineering, Xi'an Jiaotong University, Xi'an 710049, China \looseness=-2}

\author{Zuo-Guang~Ye}
\affiliation{Department of Chemistry and 4D LABS, Simon Fraser University, Burnaby, British Columbia V5A 1S6, Canada \looseness=-2}

\author{Guo-Xu~Zhang}
\affiliation{MIIT Key Laboratory of Critical Materials Technology for New Energy Conversion and Storage, School of Chemistry and Chemical Engineering, Harbin Institute of Technology, Harbin 150001, China \looseness=-2}

\author{Jingrui~Li}
\email{jingrui.li@xjtu.edu.cn}
\affiliation{State Key Laboratory for Manufacturing Systems Engineering; Electronic Materials Research Laboratory, Key Laboratory of the Ministry of Education, School of Electronic Science and Engineering, Xi'an Jiaotong University, Xi'an 710049, China \looseness=-2}

\begin{document}
\setlength{\abovedisplayskip}{2pt}
\setlength{\belowdisplayskip}{2pt}
\setlength{\abovedisplayshortskip}{2pt}
\setlength{\belowdisplayshortskip}{2pt}

\makeatletter
\def\@fnsymbol#1{\ensuremath{\ifcase#1\or \dagger\or *\or \ddagger\or
   \mathsection\or \mathparagraph\or \|\or **\or \dagger\dagger
   \or \ddagger\ddagger \else\@ctrerr\fi}}

\thispagestyle{empty}
\maketitle
\vspace{-10em}

\clearpage

\begin{abstract}
Stability is one of the key issues in mixed-halide perovskite alloys which are promising in emergent optoelectronics. Previous density-functional-theory (DFT) and machine learning studies indicate that the formation-energy convex hulls of these materials are very shallow, and stable alloy compositions are rare. In this work, we revisit this problem using DFT with special focus on the effects of configuration and vibration entropies. Allowed by the $20$-atomic models for the $\text{CsPb}(\text{I}_{1-x}^{}\text{Br}_x^{})_3^{}$ and $\text{CsPb}(\text{Br}_{1-x}^{}\text{Cl}_x^{})_3^{}$ series, the partition functions and therewith thermodynamic state functions are calculated by traversing all possible mixed-halide configurations. We can thus evaluate the temperature- and system-dependent configuration entropy, which largely corrects the conventional approach based on the ideal solution model. Finally, temperature-composition phase diagrams that include $\upalpha$, $\upbeta$, $\upgamma$, and $\updelta$ phases of both alloys are constructed based on the free energy data, for which the contribution of phonon vibrations is included.
\end{abstract}


\clearpage

\section{Introduction}

Halide perovskites ($\text{ABX}_3^{}$ with $\text{A}$ being monovalent, $\text{B}$ bivalent, and $\text{X}=\text{I},\text{Br},\text{ or Cl}$) have become a promising class of materials for emergent optoelectronics. Perovskite solar cells (PSCs) have achieved a record power conversion efficiency of $26.1\%$ \cite{NRELchartALT}, which caught up with the conventional single crystalline silicon devices that are more expensive and currently market dominating \cite{MaF2023,Szako2023}. Remarkable advances have also been seen in perovskite light-emitting diodes (PeLEDs), with high brightness, high external quantum efficiency, and excellent monochromaticity realized in devices covering a large emission wavelength range from near infrared to blue lights \cite{LuM2019,LiuXK2021,Fakharuddin2022}. Today there are still barriers on their ways to commercialization, especially the instability of these materials against environmental stresses such as heat, moisture, and oxygen \cite{HuangJ2017,ZhouY2019,ParkBw2019,HuT2021,ChenB2022}.

The many members in the $\text{ABX}_3^{}$ family, stemming from the variety of $\text{A}$, $\text{B}$, and $\text{X}$ candidates, make compositional engineering an important method to tune the properties of halide perovskites \cite{CorreaBaena2017,CorreaBaena2017b,CorreaBaena2019,Saliba2019,LuM2019}. Several strategies are widely used today. For example, halide alloying is the decisive mean to tune the perovskites band gap for different indoor light harvesting scenarios and light emission wavelengths \cite{Adjokatse2017,Chiba2018,ZhaoL2019,LuoM2020,Karlsson2021}. In PSCs, $\text{A}$-site mixing can regulate the perovskite lattice parameters \cite{CorreaBaena2017,CorreaBaena2019,SunS2021}, so that the band gap can be finely adjusted toward the optimal value according to Shockley-Queisser's detailed balance model \cite{Shockley61}. $\text{B}$-site substitution is required for the design of environmentally friendly low-lead or lead-free perovskites \cite{Swarnkar2018,deAngelis2021,WangX2023}. These strategies, however, introduce instability problems especially phase segregation into the materials \cite{LuM2019,LiuL2021,LiuXK2021}.

Computational studies, primarily using density functional theory (DFT), are playing an important role in exploring the materials stability and properties of perovskite alloys. The biggest challenge is certainly the many, in principal infinite, possible configurations of alloyed ions. For small systems, these configurations can be accessed with a traverse manner \cite{GaoW2018}. Otherwise techniques beyond DFT, such as cluster expansion and machine learning, have been employed in combination with some energy minimizing algorithms, e.g., Monte Carlo simulated annealing \cite{Yin14b,Bechtel2018,Dalpian2019,Laakso2022}. When using the minimal alloy formation energy as the criterion of thermodynamic stability, the results of these studies \cite{Yin14b,YiC2016,Bechtel2018,GaoW2018,Laakso2022} generally exhibit very shallow ($\sim$ a few $\text{meV}$ per perovskite unit) convex hulls on which only few compositions are located. This corresponds to two consequences: (a) most of the alloy compositions are unstable and will spontaneously decompose into some specific compositions, and (b) it is practically impossible to synthesize perovskite alloys of targeted compositions especially at finite temperatures, instead a mixture of many different compositions crossing the whole alloy space coexist in the product. However, experimental results do not agree with these predictions \cite{Liashenko2019,YuanS2021,LiuY2023}.

The idea of entropy-driven stability from high entropy alloys \cite{Cantor2004,Jones2014} have been adapted for non-metallic systems such as oxides \cite{Rost2015,Divilov2024} and perovskites \cite{WangX2022}. As notable examples for the entropy stability effect in halide perovskites, Yi et~al. claimed that adding $\text{Cs}$ into $\text{FAPbI}_3^{}$ ($\text{FA}$ stands for formamidinium) can suppress the formation of the undesired $\updelta$ phase \cite{YiC2016}, Gao et al. found perfect agreement between the theoretical prediction and experiment for stable $\text{FA}_{1-x}^{}\text{Cs}_x^{}\text{SnI}_3^{}$ alloy \cite{GaoW2018}, and Wang et al. found entropy effect significant in stabilizing double-perovskite alloys \cite{WangX2022}. Nevertheless, the evaluation of \textit{mixing} entropy in these works was based on the ideal solution model, i.e., all systems (configurations) equally contribute to the thermodynamic properties of the canonical ensemble. The free energy of alloy formation was then calculated by superimposing the mixing entropy term over the calculated (usually the minimal) alloy formation energy. This approach is problematic, since (a) the errors of both entropy and (internal) energy can be large, and (b) when solving a thermodynamic problem for an ensemble, it is physically improper to take the energy of one particular system (configuration) on the one hand, and use the concept of entropy that accounts for all possible configurations on the other hand.

In principle, we must have the knowledge of all configurations if we want to precisely solve the thermodynamic problem of an canonical ensemble. However, this is usually infeasible as the combinatorial number grows rapidly with the size of model system. To this end, fast (and precise) evaluation of energy of each model system would be needed, such as machine learning \cite{Laakso2022}. In this work, we study the thermodynamic properties by performing a DFT traversal of $20$-atomic models of binary mixed-halide perovskites $\text{CsPb}(\text{I}_{1-x}^{}\text{Br}_x^{})_3^{}$ and $\text{CsPb}(\text{Br}_{1-x}^{}\text{Cl}_x^{})_3^{}\,$. These all-inorganic perovskites are advantageous in materials stability and thus promising in perovskite optoelectronics \cite{Ouedraogo2020,TianJ2020,YangS2023,YuanS2021,LiuY2023,ZhouW2023}. All four commonly known phases (cubic $\upalpha$, tetragonal $\upbeta$, orthorhombic $\upgamma$, and non-perovskite $\updelta$) of $\text{CsPbX}_3^{}$ are considered. The reasons that we choose these systems are as follows. (a) The model system is not too large to traverse. For example, the number of configurations reaches its maximum $C_{12}^6=924$ at $x=0.5$, and this number can be reduced to a certain extent because of symmetry. (b) The model system is not too small to exhibit disorder. (c) There are several previous DFT-based studies of these alloy series that can serve as references to this work \cite{Yin14b,Bechtel2018,Laakso2022}. Previous DFT-based studies of $\text{Cs}$-based pure perovskites (such as $\text{CsPbI}_3^{}$ and $\text{CsPbBr}_3^{}$) indicated that the $\upalpha$ and $\upbeta$ phases are essentially dynamical mixtures of disordered structures \cite{Klarbring2019,WangX2021,LiJ2023a}. In this paper, we tackle this issue beyond pure compounds by constructing the phase diagrams for alloys. Contribution from the lattice (phonon) vibrations was taken into account for this purpose.

The remainder of this paper is organized as follows. In Sec.~\ref{sec:theory}, we briefly outline the thermodynamic theory of the perovskite alloys ensemble, the investigated systems, and computational details. In Sec.~\ref{sec:results}, the energy level distribution of each alloy and phase is analyzed. Thermodynamic state functions at $300~\text{K}$ are then calculated, from which we correctly identify the entropy-driven stability effect, and predict the alloy stability based on the Helmholtz free energy convex hull. We also compute the lattice constants and band gaps of both alloy series, and analyze how they evolve with composition. Finally, we construct the phase diagrams for both series with including the phonon vibrations into the free energy calculations. Section~\ref{sec:conclusion} concludes with a summary.

\section{Theory and computational details}\label{sec:theory}

\subsection{Thermodynamics of mixed-halide perovskites}

In this paper, the formation energy of an alloy structure is calculated from the total-energy change during the alloy formation reaction
\begin{align}
(1-x) \text{CsPbX}_3^{} + x \text{CsPbX}^{\prime}_3 &\rightarrow \text{CsPb}(\text{X}_{1-x}^{}\text{X}^{\prime}_x)_3^{}\,, \label{formation}
\end{align}
i.e.,
\begin{align}
& \Delta E(\text{CsPb}(\text{X}_{1-x}^{}\text{X}^{\prime}_x)_3^{}) = E(\text{CsPb}(\text{X}_{1-x}^{}\text{X}^{\prime}_x)_3^{}) 
- (1-x) E(\text{CsPbX}_3^{}) 
- x E(\text{CsPbX}^{\prime}_3) \label{energy}
\end{align}
with all terms in the right hand side calculated using DFT. It is then referred to as the ``energy level'' in this paper and accordingly the ``$\Delta$'' is omitted for clarity in the following if no confusion were introduced.

All thermodynamic state functions are calculated based on the energy levels. When only considering the equilibrium geometry without including phonon vibrations, the partition function of the canonical ensemble of $n$-cell systems is
\begin{align}
Z &= \sum_i g_i^{} \text{e}^{-nE_i^{}/k_{\text{B}}^{}T} \label{partition}
\end{align}
where $E_i^{}$ and $g_i^{}$ are the energy per single cell and the degree of degeneracy of the $i$th energy level, $k_{\text{B}}^{}$ the Boltzmann constant, and $T$ the temperature. The sum runs over all possible states, whose total number is $N=\sum_ig_i^{}\,$. From $Z$ we can calculate the Helmholtz free energy
\begin{align}
F &= U - TS = -\frac{k_{\text{B}}^{}T}{n} \ln(Z) \label{Helmholtz}
\end{align}
which is used as the stability criterion throughout this paper. Accordingly, the internal energy and the configuration entropy per single cell are
\begin{align}
U &= \frac{1}{Z} \sum_i g_i^{}  E_i^{}\text{e}^{-nE_i^{}/k_{\text{B}}^{}T}\,, \label{internal} \\
S &
= \frac{1}{ZT} \sum_i g_i^{}  E_i^{}\text{e}^{-nE_i^{}/k_{\text{B}}^{}T} + \frac{k_{\text{B}}^{}}{n} \ln(Z)\,, \label{entropy}
\end{align}
respectively.

\begin{landscape}
\begin{figure}[H]
\includegraphics[clip,trim=0.9in 8.0in 0.9in 0.7in]{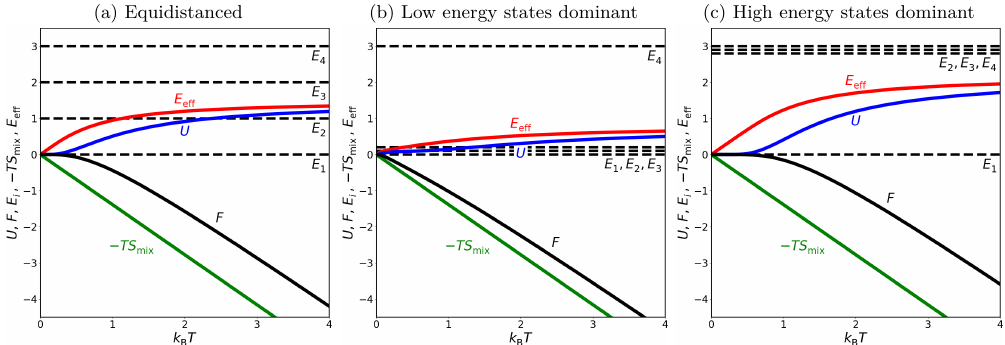}
\caption{Thermodynamic state functions $U$ (blue) and $F$ (black) of (a) an equidistanced, non-degenerate, (b) low energy levels dominating, and (c) high energy levels dominating four-level system whose levels $\{E_i^{}\vert i=1,2,3,4\}$ are marked by dashed black lines. Also shown are the entropy of mixing (in terms of $-TS_{\text{mix}}^{}\,$, green) and the effective energy $E_{\text{eff}}^{}$ (red), given by Eqs.~(\ref{entropy_old}) and (\ref{effective_energy}), respectively.}
\label{e_eff_vs_avg}
\end{figure}
\end{landscape}

In the conventional approach to perovskite alloys, the entropy of mixing (per single cell) $S_{\text{mix}}^{}$ based on the ideal solution model was generally adopted for the configuration entropy. 
$S_{\text{mix}}^{}$ is evaluated by taking the logarithm of $N$ which equals the combinatorial number $C_{3n}^{3xn}\,$:
\begin{align*}
S_{\text{mix}}^{} &= \frac{\alpha}{3n} k_{\text{B}}^{} \ln(N) 
= -\alpha k_{\text{B}}^{} [ x \ln(x) + (1-x) \ln(1-x) ]
\end{align*}
with the coefficient $\alpha$ to determine. Let he partition function have the form
\begin{align}
Z &= N \text{e}^{-nE_{\text{eff}}^{}/k_{\text{B}}^{}T}\,,
\label{partition_old}
\end{align}
we can rewrite Eq.~(\ref{Helmholtz}) into the following form
\begin{align}
F 
&= -\frac{k_{\text{B}}^{}T}{n} \ln\Big(N^{\alpha/3}\text{e}^{-nE_{\text{eff}}^{}/k_{\text{B}}^{}T}
\Big)\,. \nonumber
\end{align}
We thus have $\alpha=3$ and
\begin{align}
S_{\text{mix}}^{} &= -3k_{\text{B}}^{} [ x \ln(x) + (1-x) \ln(1-x) ]\,. \label{entropy_old}
\end{align}
This is easy to understand as the contribution comes from three halide ions in a single cell. Eq.~(\ref{partition_old}) describes an ensemble in which all $N$ mixed-halide configurations have the same temperature-dependent effective energy
\begin{align}
E_{\text{eff}}^{} &= -\frac{k_{\text{B}}^{}T}{n} [ \ln(Z) - \ln(N) ]\,. \label{effective_energy}
\end{align}
$E_{\text{eff}}^{}$ is not equal to the canonical ensemble mean energy $E$ calculated by Eq.~(\ref{internal}) unless $T=0$ or all energy levels are degenerate. As an example, Fig.~\ref{e_eff_vs_avg} shows the temperature-dependence of $F$, $U$, and $E_{\text{eff}}^{}$ of three different four-level systems. For an equidistanced, non-degenerate four-level system [Fig.~\ref{e_eff_vs_avg}(a)], $E_{\text{eff}}^{}$ obviously differs from $E_{\min}^{}=0$ once $T$ becomes positive. As a result, the Helmholtz free energy evaluated by $(E_{\min}^{}-TS_{\text{mix}}^{})$ deviates from the correct value $F=E_{\text{eff}}^{}-TS_{\text{mix}}^{}\,$. At high temperatures (e.g., when $k_{\text{B}}^{}T$ is significantly larger than the energy level difference), their difference becomes almost constant, as the energy levels become nearly equally distributed so that $E_{\text{eff}}^{}$ approaches their arithmetic mean. As expected, the difference between $F$ and $(E_{\min}^{}-TS_{\text{mix}}^{})$ is smaller (larger) for multiple-level systems in which low energy levels dominate [Fig.~\ref{e_eff_vs_avg}(b)], and larger for high-level dominant systems. Figure~\ref{e_eff_vs_avg} also shows that the more high-level dominant, the higher the temperature at which the entropy effect can be obviously observed. This is natural as the contribution of high energy levels can only be large at high enough temperatures.

Because of the lack in traversing the configuration space or in approximating the density of states (DOS), neither the canonical ensemble mean energy $U$ given by Eq.~(\ref{internal}) nor the effective energy by Eq.~(\ref{effective_energy}) could be evaluated in previous studies. Instead, the minimal energy $E_{\min}^{}$ of all sampled configurations was usually chosen, over which the $-TS_{\text{mix}}^{}$ term with $S_{\text{mix}}^{}$ given by Eq.~(\ref{entropy_old}) was superimposed to calculate the Helmholtz free energy. This approach is for sure physically incorrect as it always results in a too low Helmholtz free energy, i.e., it effectively overestimates the entropy-driven stabilization effect. This can be rationalized by an obvious contradiction: $E_{\min}^{}$ corresponds to only a few (or even one) configurations, while with $-TS_{\text{mix}}^{}$ all configurations are regarded having equal population. For a correct understanding of the thermodynamics of mixed-halide perovskite alloys, we perform a traverse of all possible configurations at each composition in this paper which is allowed by the relatively small model systems.


\subsection{Alloy configurations of the four phases}

\begin{figure}[H]
\includegraphics[clip,trim=1.2in 8.5in 1.2in 0.7in]{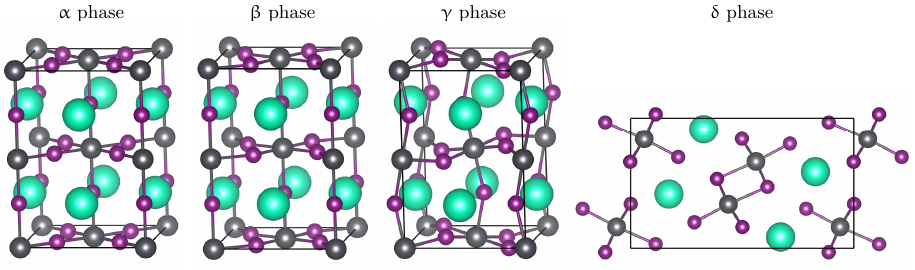}
\caption{The $20$-atomic model systems of all four investigated phases. $\text{Cs}$, $\text{Pb}$, and halides are colored in green, dark gray, and purple, respectively.}\label{model}
\end{figure}

We chose a $\,\sqrt[]{2}\times\,\sqrt[]{2}\times2$ ($20$ atoms) model to represent the $\upalpha$ (space group $Pm\bar{3}m$), $\upbeta$ ($P4/mbm$), and $\upgamma$ ($Pnma$) phases , as it is the smallest model for the structure of $\upgamma$ phase. The primitive cell of the non-perovskite $\updelta$ phase ($Pnma$) contains also $20$ atoms. Model systems are shown in Fig.~\ref{model}.

Symmetry analysis was performed to identify all non-equivalent structures and their degrees of degeneracy at each possible composition. Table~\ref{degeneracy} shows that the degree of degeneracy increases from $\upgamma$ to $\upbeta$, i.e., some non-equivalent structures of the former become equivalent in the latter. This is because of the higher symmetry of $P4/mbm$ than $Pnma$. Similar trend is found between the $P4/mbm$ and $Pm\bar{3}m$ phases for compositions $x=\frac{2}{12},\frac{4}{12},\frac{6}{12}$. For other compositions, the number and the degrees of degeneracy of non-equivalent structures of $\upalpha$ and $\upbeta$ phases are equal to each other although the symmetry of the former is higher. Detailed analysis are available in Sec.~S1 of Supporting Information (SI).

Table~\ref{degeneracy} shows that the number of non-equivalent structures of $\updelta$ phase is larger than any of other phases at each composition, which indicates the lowest symmetry of this phase despite the same space group as for the $\upgamma$ phase. This is primarily due to that there are three different types of halide sites in the $\updelta$ phase (see SI, Fig.~S6), compared to two in $\upgamma$. As a result of the four halides of each type within the primitive cell model, most of the non-equivalent alloy structures of $\updelta$ phase are four-fold degenerate.

We must note that, strictly speaking, the terms ``orthorhombic'', ``tetragonal'', and ``cubic'' are no longer strictly available for alloys, as halide mixing changes the symmetry of the whole system. Nevertheless, we still use them to name the phases in this work, and correspondingly applied constraints to the atomic structure in DFT calculations. Specifically, for the orthorhombic ($\upgamma$ and $\updelta$) phases we only forced the three lattice vectors orthogonal to each other with no constraints on the atomic positions. For the tetragonal ($\upbeta$) phase, we further ensured $a=b$, and all $\text{Cs}^+$, $\text{Pb}^{2+}$, and out-of-plane $\text{X}^-$ ions at their Wyckoff positions as in a standard $P4/mbm$ structure, while the in-plane $\text{X}^-$ ions were allowed to move within the $ab$ planes. For the cubic ($\upalpha$) phase, the lattice vectors are forced to maintain the perfect cubic lattice and all ions are frozen in the corresponding Wyckoff positions of $Pm\bar{3}m$.

\begin{landscape}
\begin{table}[H]
\caption{Degree of degeneracy of each composition (in terms of the number ratio of included halide ions in the model) of all four phases. As an example, $3\times2$ indicates $3$ non-equivalent mixed-halide alignments each characterized by degree of degeneracy $2$.}
\label{degeneracy}
\begin{tabular}{c@{\hspace{2em}}l@{\hspace{2em}}l@{\hspace{2em}}l@{\hspace{2em}}l} \hline\hline
${\color{white}{^{\prime}}}\text{X}:\text{X}^{\prime}$ & \multicolumn{1}{c}{$\upalpha$} & \multicolumn{1}{c}{$\upbeta$} & \multicolumn{1}{c}{$\upgamma$} & \multicolumn{1}{c}{$\updelta$} \\ \hline
${\color{white}{0}}1:11$  & $4$, $8$ & $4$, $8$ & $4$, $8$ & $3\times4$ \\
${\color{white}{0}}2:10$  & $2\times2$, $2\times4$, $6$, $2\times8$, $32$ & $3\times2$, $3\times4$, $2\times8$, $32$ & $3\times2$, $7\times4$, $32$ & $9\times2$, $12\times4$ \\
${\color{white}{0}}3:{\color{white}{0}}9$ & $4$, $3\times8$, $8\times16$, $2\times32$ & $4$, $3\times8$, $8\times16$, $2\times32$ & $4$, $7\times8$, $10\times16$ & $55\times4$ \\
${\color{white}{0}}4:{\color{white}{0}}8$ & $2\times2$, $3$, $4$, $11\times8$, $12$ & $1$, $3\times2$, $2\times4$, $12\times8$, & $1$, $7\times2$, $28\times8$, $8\times32$ & $3\times1$, $18\times2$, $114\times4$ \\
& $8\times16$, $4\times32$, $2\times64$ & $8\times16$, $4\times32$, $2\times64$ & & \\
${\color{white}{0}}5:{\color{white}{0}}7$ & $7\times8$, $16\times16$, $11\times32$, & $7\times8$, $16\times16$, $11\times32$, & $15\times8$, $28\times16$, $7\times32$ & $198\times4$ \\
& $2\times64$ & $2\times64$ & & \\
${\color{white}{0}}6:{\color{white}{0}}6$ & $12\times4$, $10\times8$, $12$, $9\times16$, & $15\times4$, $10\times8$, $9\times16$, & $35\times4$, $21\times16$, $14\times32$ & $30\times2$, $204\times4$, $6\times8$ \\
& $12\times32$, $4\times64$ & $12\times32$, $4\times64$ & & \\ \hline\hline
\end{tabular}
\end{table}
\end{landscape}


\subsection{Effect of phonon vibration}

Within the harmonic approximation, the $k_j^{}$th vibrational energy levels of the $j$th phonon mode of an arbitrary structure is
\begin{align}
\varepsilon_{jk}^{} &= \bigg(k_j^{}+\frac{1}{2}\bigg)\hbar\omega_j^{}\,, \quad k=0,1,\ldots
\end{align}
where $\hbar$ is the reduced Planck constant and $\omega$ the frequency. Thus, the overall (total plus vibrational) energy levels of the $i$th configuration in the canonical ensemble are
\begin{align}
E_{i\bm{k}_i^{}}^{} &= E_i^{(0)} + \sum_{j_i^{}} \varepsilon_{j_i^{}k_{ij_i^{}}^{}}^{} = E_i^{(0)} + \sum_{j_i^{}} \bigg(k_{ij_i^{}}^{}+\frac{1}{2}\bigg)\hbar\omega_{ij_i^{}}^{}\,,
\end{align}
where $E_i^{(0)}$ is the (e.g., DFT-calculated) energy at the equilibrium position as $E_i^{}$ in Eq.~(\ref{partition}), $\bm{k}_i^{}=\{k_{i1}^{},k_{i2}^{},\ldots\}$ is a compact notation for the vibrational quantum numbers with $k_{ij_i^{}}^{}=0,1,2,\cdots$ denoting the vibrational energy level of the $j_i^{}$th phonon mode. The index $i$ in the subscript of $j$, $\omega$, and $k$ distinguishes the vibrational spectra of different configurations. The total partition function then reads
\begin{align}
Z &= \sum_{i\bm{k}_i^{}} g_i^{} \text{e}^{-nE_{i\bm{k}_i^{}}^{}/k_{\text{B}}^{}T} \nonumber \\
&= \sum_i g_i^{} \text{e}^{-n\big(E_i^{(0)}+\text{ZPE}_i^{}\big)/k_{\text{B}}^{}T} \prod_{j_i^{}} \Bigg(\frac{1}{1-\text{e}^{-\hbar\omega_{ij_i^{}}^{}/k_{\text{B}}^{}T}}\Bigg)^n \nonumber \\
&= \sum_i g_i^{} \text{e}^{-n\big(E_i^{(0)}+\text{ZPE}_i^{}-TS_i^{\text{phonon}}\big)/k_{\text{B}}^{}T} \label{partition_total}
\end{align}
where for the $i$th configuration, the zero point energy is
\begin{align}
\text{ZPE}_i^{} = \sum_j\frac{1}{2}\hbar\omega_{ij}^{}\,, \label{zpe}
\end{align}
and the phonon entropy
\begin{align}
S_i^{\text{phonon}} &= -k_{\text{B}}^{} \sum_j \ln\Big(1-\text{e}^{-\hbar\omega_{ij}^{}/k_{\text{B}}^{}T}\Big)\,. \label{single_phonon_entropy}
\end{align}
Both can be obtained from a regular phonon calculation for each individual structure (configuration). The Helmholtz free energy can be calculated by inserting Eqs.~(\ref{partition_total}\--\ref{single_phonon_entropy}) into Eq.~(\ref{Helmholtz}). Effectively, the contribution of all vibrational energy levels $\{E_{i\bm{k}_i^{}}^{}\}$ of the $i$th structure can be regarded as a ``correction'' to the DFT calculated energy of its equilibrium geometry, $E_i^{(0)}$, by the ZPE and phonon entropy.

\subsection{Computational details}

For DFT structure optimization calculations, we chose the Perdew-Burke-Ernzerhof exchange-correlation functional for solids (PBEsol) \cite{Perdew2008} implemented in the all-electron numeric-atom-centered orbital code FHI-aims \cite{Blum09,HavuV09,RenX12,Levchenko15}, as PBEsol describes the lattice constants of halide perovskites well with moderate computational cost \cite{YangRX2017,Bokdam2017,Seidu2021a,LiJ2023a}. 
Scalar relativistic effects were included by means of the zero-order regular approximation \cite{vanLenthe93}, while spin-orbit coupling (SOC) was further included in PBEsol0 calculations. Standard FHI-aims tier-2 basis sets were used in combination with a $\Gamma$-centered $6\times6\times4$ ($10\times5\times3$ for $\updelta$) $k$-point mesh.

We used the analytical stress tensor \cite{Knuth15} implemented in FHI-aims for the lattice constant optimization of the orthorhombic phase structures. For the tetragonal structures, the lattice constants $a=b$ and $c$ were optimized using the recently developed Bayesian Optimisation Structure Search (BOSS) package, a machine-learning-based structure search scheme for accelerated and unbiased potential-energy-surfaces computation \cite{Todorovic2019}. For such a two-dimensional optimization problem, BOSS used the surrogate model that was fitted to all existing data to determine the $(a,c)$ coordinates of the next point to sample. Then DFT geometry optimization was performed without relaxing the lattice constants (many atomic coordinates were frozen, too, as already alluded to). The data base was updated by returning the energy of the relaxed structure to BOSS, and such an iteration was performed till convergence. The approach to the cubic phase is similar, in which BOSS optimized $a=b=c$ by handling a one-dimensional problem. As all atoms were fixed at their Wyckoff positions, only single-point DFT calculations were carried out during each BOSS iteration.

Phonon properties were computed using a finite displacement approach implemented in the Phonopy program \cite{Togo2015,Togo2023}. Based on convergence test calculations, displacement of $0.01~\text{\AA}$ and tier-1 basis sets were chosen to calculate the atomic forces and thus the Hessian matrix. 

The results of all DFT calculations of the perovskite alloys are available from the NOMAD (Novel Materials Discovery) repository. \cite{NoMaD-mixX_z12}.

\section{Results and discussion}\label{sec:results}

\subsection{Energy spectra of each phase}

\begin{figure}[H]
\includegraphics[clip,trim=1.4in 2.8in 1.5in 0.7in]{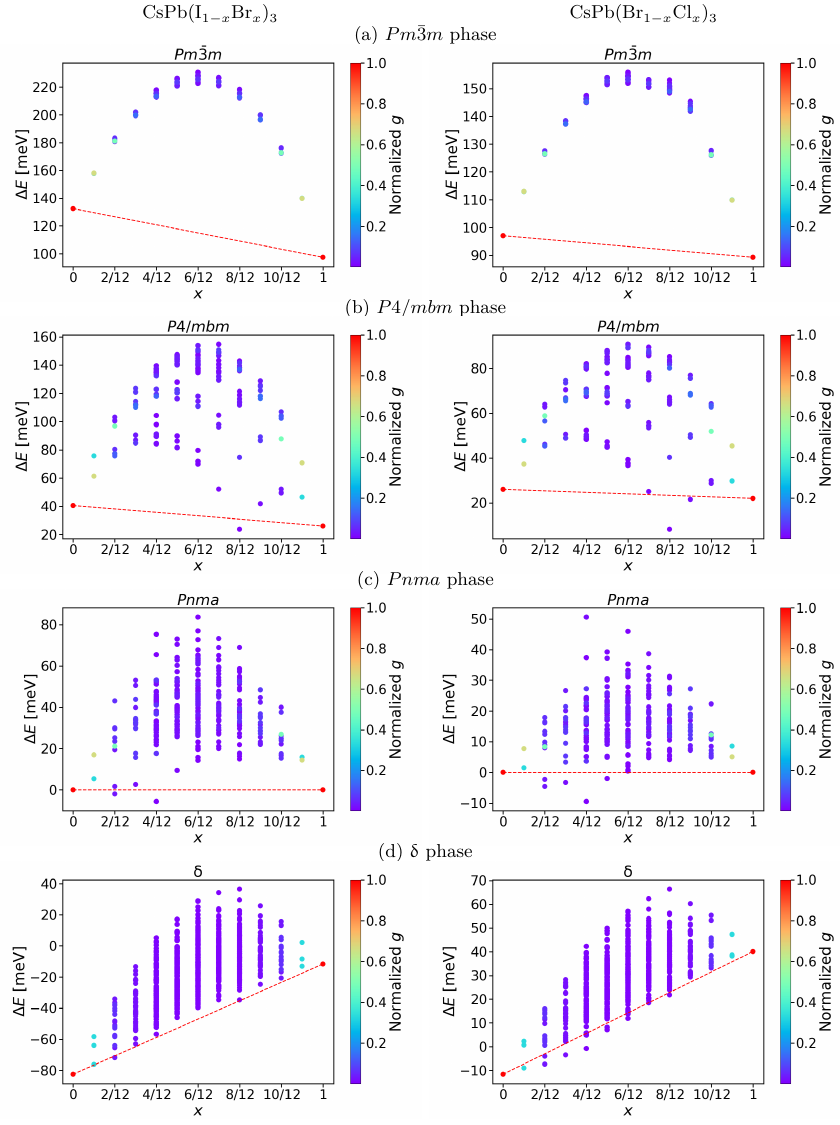}
\caption{DFT calculated formation energies (in $\text{meV}$ per perovskite unit) of mixed-halide perovskite alloys $\text{CsPb}(\text{I}_{1-x}^{}\text{Br}_x^{})_3^{}$ (left) and $\text{CsPb}(\text{Br}_{1-x}^{}\text{Cl}_x^{})_3^{}$ (right) in all four phases, (a) $\upalpha$ ($Pm\bar{3}m$), (b) $\upbeta$ ($P4/mbm$), (c) $\upgamma$ ($Pnma$), and (d) $\updelta$ ($Pnma$), all compositions, and all configurations. The color of each scattered data point indicates the normalized degeneracy.}
\label{energies}
\end{figure}

Figure~\ref{energies} shows the formation energies of all considered alloy phases, compositions, and configurations. 
All data are calculated using Eq.~(\ref{energy}) with taking the DFT total energies of pure perovskites [$\text{X}=\text{I}$, $\text{X}^{\prime}=\text{Br}$ for $\text{CsPb}(\text{I}_{1-x}^{}\text{Br}_x^{})_3^{}\,$, $\text{X}=\text{Br}$, $\text{X}^{\prime}=\text{Cl}$ for $\text{CsPb}(\text{Br}_{1-x}^{}\text{Cl}_x^{})_3^{}$] in the $Pnma$ phase. In general, these two alloy series exhibit similar energetic characters.

Representative structures at each $x$ in each phase, e.g., with the lowest and highest formation energies, are provided in SI, Sec.~S2. 
The following discussion mainly focuses on the DOS which can be derived from the energy spectra
\begin{align}
\rho(E) &= A \sum_i g_i \text{e}^{-(E-nE_i^{})^2/\sigma^2}
\end{align}
with $A$ the normalization factor and $\sigma$ the width parameter. In this paper we chose $\sigma=5~\text{meV}$ which can properly balance the global and local information in the DOS. With $\rho(E)$, the partition function in Eq.~(\ref{partition}) can be rewritten in the integral form as
\begin{align}
Z &= \int \rho(E) \text{e}^{-E/k_{\text{B}}^{}T} \operatorname{d}E\,.
\end{align}

\subsubsection{\texorpdfstring{$Pm\bar{3}m$}{} phase}

\begin{figure}
\includegraphics[clip,trim=1.8in 2.8in 1.5in 0.7in]{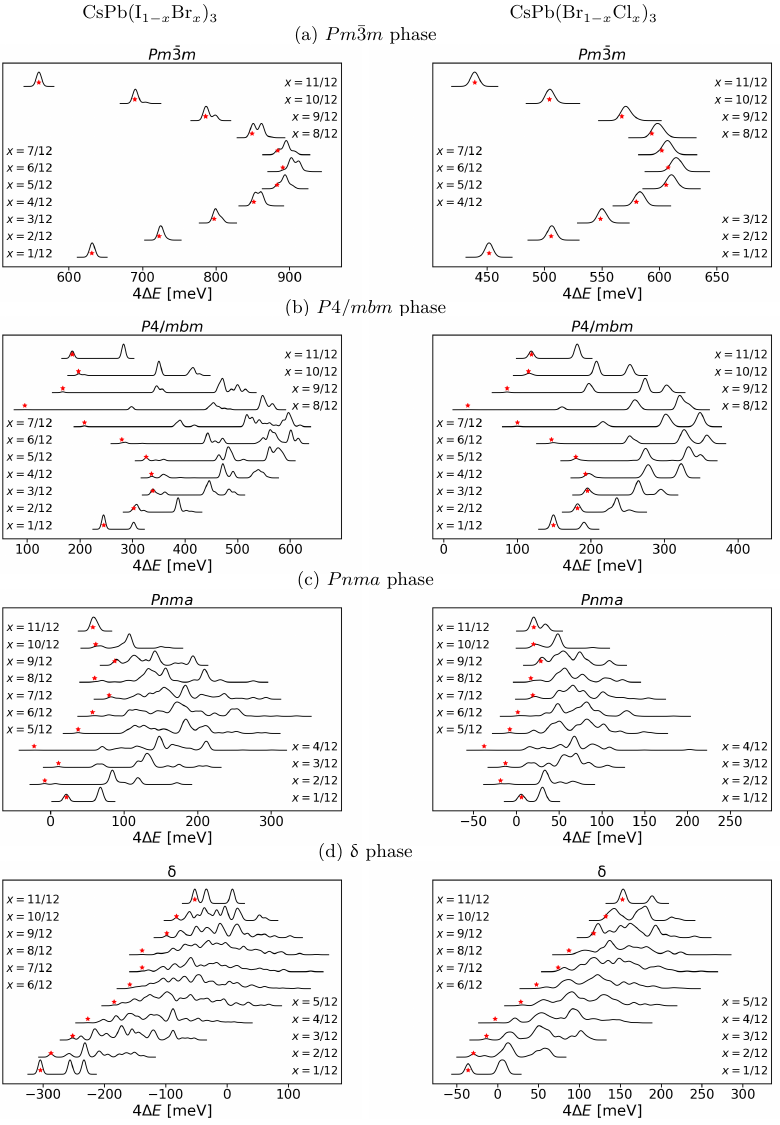}
\caption{DFT calculated energy-level distribution (i.e., unnormalized DOS) of mixed-halide perovskite alloys $\text{CsPb}(\text{I}_{1-x}^{}\text{Br}_x^{})_3^{}$ (left) and $\text{CsPb}(\text{Br}_{1-x}^{}\text{Cl}_x^{})_3^{}$ (right) in all four phases, (a) $\upalpha$ ($Pm\bar{3}m$), (b) $\upbeta$ ($P4/mbm$), (c) $\upgamma$ ($Pnma$), and (d) $\updelta$ ($Pnma$), all compositions, and all configurations. The red star in each plot indicates the lowest energy level of the particular phase. Note the energies are given in $\text{meV}$ per model system, i.e., $4$ perovskite units.}
\label{dos}
\end{figure}

For the cubic $\upalpha$ phase, the energy data points of both alloy series [Fig.~\ref{energies}(a)] exhibit similar features. First, no convex hull can be drawn as all $x\in(0,1)$ data points are well above the line connecting the two pure-compound data points. The data points actually rather show a typical concave character. Second, at each composition, the alloy formation energies of either alloy series spread over a relatively small range ($<9~\text{meV}$), which can be obviously observed in the energy-level distribution function, i.e., DOS [Fig.~\ref{dos}(a)]. This signifies that in terms of the cubic lattice with applying the aforementioned constrains on the lattice constants and ionic positions, the dependence of the model's total energy on the mixed ion configuration is rather weak.

From these data we can anticipate that in the cubic phase, the investigated alloys are not stable at low temperatures as they have strong tendency to decompose into pure perovskite compounds. Because of the narrow range of energy-level distribution, the entropy effect could be large and close to evaluated based on ideal solution model [cf. Fig.~\ref{e_eff_vs_avg}(b)]. Nevertheless, high temperatures would be required to stabilize the alloys.

\subsubsection{\texorpdfstring{$P4/mbm$}{} phase}

Data points with negative formation energies within this phase, i.e., below the red dashed lines in Fig.~\ref{energies}(b), are observed for both series. For $\text{CsPb}(\text{I}_{1-x}^{}\text{Br}_x^{})_3^{}\,$, there is only one such configuration at $x=\frac{8}{12}$ (i.e., $\text{CsPbIBr}_2^{}$) whose degree of degeneracy is $1$ meaning a highly ordered mixed-halide alignment (all of the larger halide anions, here $\text{I}^-$, are located out-of-plane). Similar configuration contributes to the lowest-energy data point within $\text{CsPb}(\text{Br}_{1-x}^{}\text{Cl}_x^{})_3^{}\,$, too; yet another negative-formation-energy data point can be found at $x=\frac{9}{12}$ (also with a highly ordered alignment). All other data points are well above the red dashed lines in Fig.~\ref{energies}(b) indicating large positive formation energies.

Figure~\ref{dos}(b) shows that the contribution of above mentioned lowest energy data [marked by red stars at $x=\frac{8}{12}$ for $\text{CsPb}(\text{I}_{1-x}^{}\text{Br}_x^{})_3^{}\,$, and $x=\frac{8}{12},\frac{9}{12}$ for $\text{CsPb}(\text{Br}_{1-x}^{}\text{Cl}_x^{})_3^{}$] to the thermodynamic properties of the ensemble is negligible. Actually, at most of the compositions $x$ for both series, the energy-level distribution of the $P4/mbm$ phase exhibits a high energy level dominant character. As indicated by Fig.~\ref{e_eff_vs_avg}(c), such a character corresponds to a large overestimation of the entropy effect by the traditional model.

\subsubsection{\texorpdfstring{$Pnma$}{} (\texorpdfstring{$\upgamma$}{}) phase}

For this most stable perovskite phase of all-inorganic perovskites, our DFT data [Fig.~\ref{energies}(c)] exhibit similar profile as previously calculated by DFT and ML \cite{Laakso2022} and by DFT and CE \cite{Yin14b,Bechtel2018} except that the energy ranges in our results are obviously larger, because the traversing approach forces to explore the high energy regions. For both alloys, shallow ($<10~\text{meV}$ per perovskite unit) convex hulls are observed which contain very few negative data points. 
Stable structures are found at $x=\frac{1}{6}$ and $\frac{1}{3}$, both exhibiting highly ordered mixed-halide alignments thus being associated with low degrees of degeneracy ($2$ and $1$, respectively). Configurations with negative formation energies at $x=\frac{1}{2}$ and $\frac{2}{3}$ are not found for $\text{CsPb}(\text{Br}_{1-x}^{}\text{Cl}_x^{})_3^{}$ as in Ref.~\cite{Laakso2022}, naturally due to the smaller computational model in this paper.

As illustrated by Fig.~\ref{dos}(c), at most of the $x$ values of both alloys, the DOS is distributed within a large range of energy which is well above the minimum. Thus, the high level dominant model [Fig.~\ref{e_eff_vs_avg}(a)] is proper to describe these ensembles, yet the energy difference between the minimum and the major distributed region is not as large as in $P4/mbm$.

\subsubsection{\texorpdfstring{$\updelta$}{} phase}

The formation of non-perovskite $\updelta$ phase is one of the major challenge in all-inorganic perovskite solar cells as it is photovoltaic inactive. It is the most stable polymorph of $\text{CsPbI}_3^{}\,$; for $\text{CsPbBr}_3^{}$ it is described as ``metastable'' at room temperature \cite{Marstrander1966,Aebli2020}; while for the chloride analog, this phase is experimentally not yet reported to our best knowledge. Our DFT (PBEsol) results agree well with this trend: the total energy of $\updelta$-$\text{CsPbI}_3^{}$ is lower than its $\upgamma$ counterpart by $82.5~\text{meV}$ per perovskite unit, this difference becomes much smaller ($11.7~\text{meV}$) for $\text{CsPbBr}_3^{}\,$, while the $\upgamma$ phase is more stable (by $40.1~\text{meV}$) for $\text{CsPbCl}_3^{}\,$. Note: previous report that $\upgamma$-$\text{CsPbBr}_3^{}$ is slightly more stable than $\updelta$-$\text{CsPbBr}_3^{}$ does not signify that the quality of our DFT calculations is questionable. First of all, the reverse of $\updelta$ vs. $\upgamma$ phase along $\text{I}$\--$\text{Br}$\--$\text{Cl}$ is properly produced by our calculations which is of the most importance. Second, the calculated energy difference is small, indicating the co-existence of both polymorphs at finite temperatures as experimentally revealed \cite{Marstrander1966,JungJ2023}. In addition, low-temperature measurement of the relative stability of these two $\text{CsPbBr}_3^{}$ is lacking.

For $\updelta$-$\text{CsPb}(\text{I}_{1-x}^{}\text{Br}_x^{})_3^{}$ [Fig.~\ref{energies}(d)-left], the minimum energy data point at each composition is very close to the straight line linking the two pure compounds, $\text{CsPbI}_3^{}$ and $\text{CsPbBr}_3^{}\,$. 
For $\updelta$-$\text{CsPb}(\text{Br}_{1-x}^{}\text{Cl}_x^{})_3^{}$ [Fig.~\ref{energies}(d)-right], there are few data points below this line by a few $\text{meV}$ per perovskite unit.

For most of the compositions of this phase, data points with different mixed-halide configurations are distributed over a large range of energies in a relatively uniform manner, as shown in Fig.~\ref{energies}(d) and Fig.~\ref{dos}(d). In this regard, Fig.~\ref{e_eff_vs_avg}(a) is a proper model to predict the thermodynamic properties.

\subsection{Thermodynamic state functions and the temperature effects}

\begin{figure}[H]
\includegraphics[clip,trim=1.4in 2.8in 1.5in 0.7in]{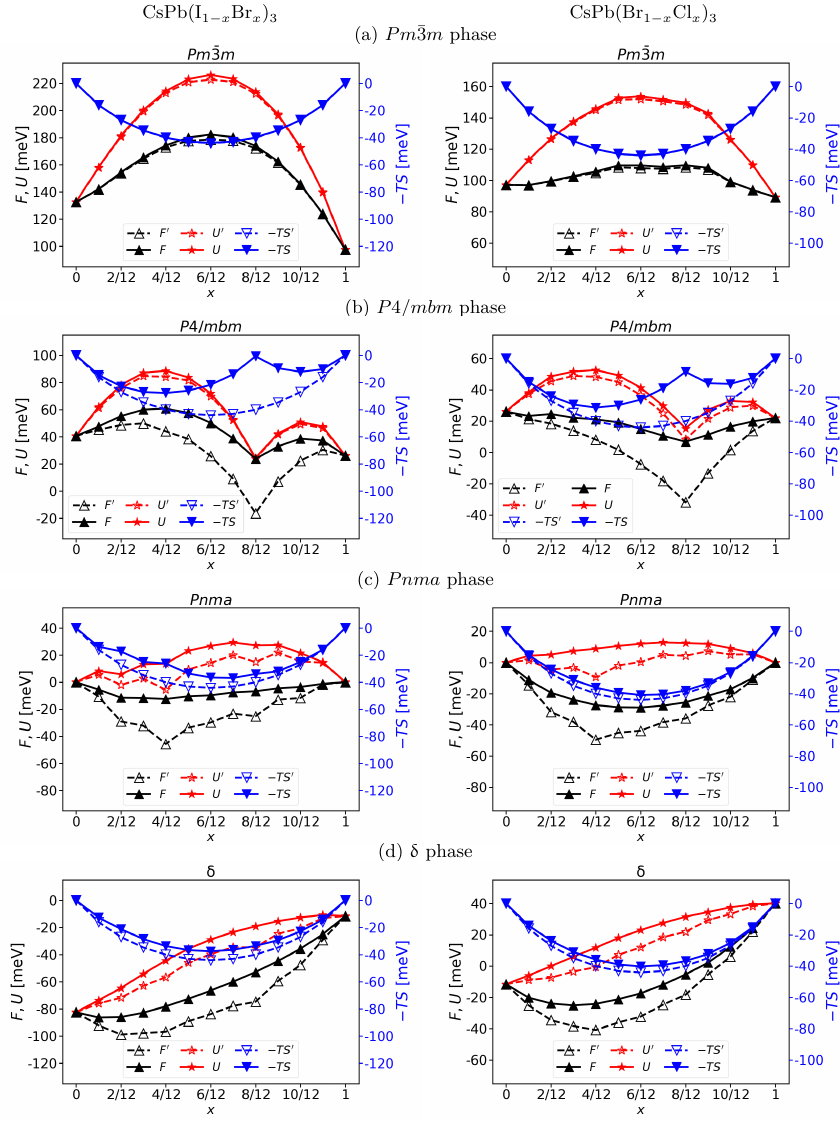}
\caption{Thermodynamic state functions $F$ (in black solid lines and upper triangles), $U$ (in red solid lines and stars), and $S$ (right $y$ axes, in blue solid lines and lower triangles, given in $-TS$) of mixed-halide perovskite alloys $\text{CsPb}(\text{I}_{1-x}^{}\text{Br}_x^{})_3^{}$ (left) and $\text{CsPb}(\text{Br}_{1-x}^{}\text{Cl}_x^{})_3^{}$ (right) in all four phases, (a) $\upalpha$ ($Pm\bar{3}m$), (b) $\upbeta$ ($P4/mbm$), (c) $\upgamma$ ($Pnma$), and (d) $\updelta$ ($Pnma$), all compositions, and all configurations at $300~\text{K}$. Also given are the functions calculated by the traditional approach, $F^{\prime}$, $U^{\prime}$, and $-TS^{\prime}$ (in the corresponding colors, dashed lines, and empty markers) calculated with Eq.~(\ref{traditional}). All observables are given in their values per perovskite unit. In each alloy, the left and right $y$ axes of all phases are with the same scale.}
\label{300k}
\end{figure}

Using Eqs.~(\ref{partition})-(\ref{entropy}) with the DFT-calculated energy levels $\{E_i^{}\}$, we calculated the thermodynamic state functions $F$ (free energy), $U$ (internal energy), and $S$ (entropy) at $T=300~\text{K}$ as illustrated in Fig.~\ref{300k}. For comparison, also shown in Fig.~\ref{300k} are the thermodynamic state functions calculated with the ``traditional approach'':
\begin{align}
\left.\begin{array}{rcl} U^{\prime} & \!\!=\!\! & E_{\min}^{} \cr
S^{\prime} & \!\!=\!\! & S_{\text{mix}}^{} \cr
F^{\prime} & \!\!=\!\! & U^{\prime} - TS^{\prime} \end{array}\right\}
\label{traditional}
\end{align}
with $S_{\text{mix}}^{}$ given by Eq.~(\ref{entropy_old}).

For the $Pm\bar{3}m$ phase [Fig.~\ref{300k}(a)], the curves of our approach and the traditional approach nearly overlap, indicating that the traditional approach is a very good approximation as already anticipated in the discussion of energy level distribution. The major difference between these two alloys is that the free energy curve of $\text{CsPb}(\text{I}_{1-x}^{}\text{Br}_x^{})_3^{}$ shows a typical concave character at $300~\text{K}$, while $F(300~\text{K})$ is relatively flat vs. $x$ in $\text{CsPb}(\text{Br}_{1-x}^{}\text{Cl}_x^{})_3^{}\,$. This is because the entropy effect of the former is not large enough to compensate the energy loss upon the halide mixing at $300~\text{K}$.

The $P4/mbm$ phases [Fig.~\ref{300k}(b)] show an obviously different character when compared to $Pm\bar{3}m$, which is certainly due to the much more complicated energy level distribution of the former. In both alloys, the traditional approach significantly underestimates the free energy ($F^{\prime}$ vs. $F$), thus overestimating the entropy stabilization effect. Further analysis shows that, at $300~\text{K}$, the difference between this and the traditional approach is small for internal energy [$<5~\text{meV}$ for $\text{CsPb}(\text{I}_{1-x}^{}\text{Br}_x^{})_3^{}$ and $<8~\text{meV}$ for $\text{CsPb}(\text{Br}_{1-x}^{}\text{Cl}_x^{})_3^{}$]. The underestimation of $F$ by the traditional approach is mainly contributed from the entropy calculation which uses the ideal solution model. It is noteworthy that these phenomena change when temperature increases to $400~\text{K}$ (cf. SI, Sec.~S4), where the $U$ and $U^{\prime}=E_{\min}^{}$ difference becomes obvious for $\text{CsPb}(\text{Br}_{1-x}^{}\text{Cl}_x^{})_3^{}$ (e.g., $>20~\text{meV}$ at $x=\frac{8}{12}$) but remains small ($<8~\text{meV}$) for $\text{CsPb}(\text{I}_{1-x}^{}\text{Br}_x^{})_3^{}\,$. This is because the energy difference between the high energy configurations and the low energy ones for $\text{CsPb}(\text{Br}_{1-x}^{}\text{Cl}_x^{})_3^{}$ is clearly smaller than for $\text{CsPb}(\text{I}_{1-x}^{}\text{Br}_x^{})_3^{}\,$, so that the high energy configurations of the former can have more significant contribution to the ensemble when temperature increases.

\begin{figure}[H]
\includegraphics[clip,trim=1.5in 8.4in 1.8in 0.7in]{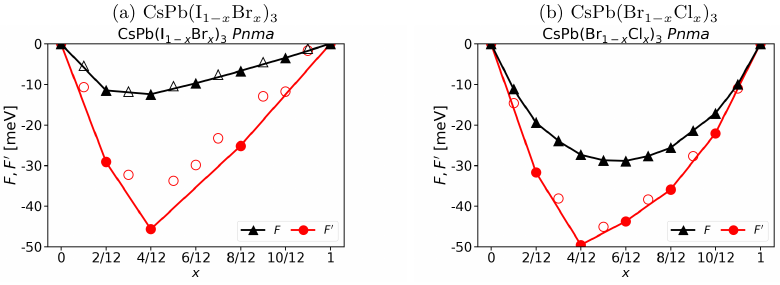}
\caption{Convex hulls of $Pnma$ phase of (a) $\text{CsPb}(\text{I}_{1-x}^{}\text{Br}_x^{})_3^{}$ and (b) $\text{CsPb}(\text{Br}_{1-x}^{}\text{Cl}_x^{})_3^{}$ at $300~\text{K}$ evaluated from the thermodynamic ensemble ($F$, in black upper triangles) and by the traditional approach ($F^{\prime}$, in red circles). Solid and empty markers indicate the free energy data on and above the hulls, respectively.}
\label{hull300k}
\end{figure}

The room-temperature thermal stability for $Pnma$, the perovskite phase with the lowest formation (internal) energy, is of particular interest. As shown by the red curves in Fig.~\ref{hull300k}, a $\sim50~\text{meV}$ deep convex hull can be expected for both alloys by superimposing the ideal-solution mixing entropy term over the minimum formation energy. In addition, only data of a few concentrations are on these hulls, indicating the alloys of other concentrations spontaneously decompose at $300~\text{K}$. However, this traditional approach obviously overestimates the stabilization effect, as shown by the much shallower hull of $F$ in Fig.~\ref{hull300k}. More specifically, the hull of $Pnma$ $\text{CsPb}(\text{I}_{1-x}^{}\text{Br}_x^{})_3^{}$ exhibits three major characters. First, it is only $12.5~\text{meV}$ deep. Second, only data $x=\frac{n}{12}$ with even $n$ are on the hull, while the others are slightly above the hull. Third, the hull between $x=\frac{4}{12}$ and $x=1$ is almost linear, thus for any alloy with composition within this part, the energy loss upon the decomposition into $x=\frac{1}{12}$ and $x=1$ species is very small, or in other words, the stabilization effect due to alloying is small. In contrast, the hull of $Pnma$ $\text{CsPb}(\text{Br}_{1-x}^{}\text{Cl}_x^{})_3^{}$ is obviously deeper ($28.9~\text{meV}$). All calculated $F$ data are found on the hull, which exhibit a smooth convex character thus indicating more pronounced stabilization effect.

Finally we take a look at the $\updelta$ phase. Figure~\ref{300k}(d) shows that these two alloys exhibit similar thermodynamic characters in this phase at $300~\text{K}$. Similar to $Pnma$, the traditional approach notably overestimates the entropy stabilization effect, while our approach results in shallower convex hulls which contain all calculated $F$ data points.

\subsection{Ensemble average of materials properties}

With the partition function $Z$ calculated by Eq.~(\ref{partition}), we can evaluate the ensemble average of materials properties ($P$) at finite temperatures by
\begin{align}
P &= \frac{1}{Z} \sum_i g_i^{}  P_i^{} \text{e}^{-nE_i^{}/k_{\text{B}}^{}T}
\end{align}
with $P_i^{}$ denoting the value of $P$ of the $i$th system.

We first discuss the effective lattice constants of the perovskite phases. They are defined as follows:
\begin{itemize}
  \item For the cubic $\upalpha$ phase, the lattice constants of our $20$-atomic model are $(\sqrt[]{2}a,\sqrt[]{2}a,2a)$, thus the effective lattice constant for the cubic single (i.e., $5$-atomic) cell is $a$. The effective single-cell volume, i.e., the volume occupied by one $\text{CsPbX}_3^{}$ perovskite unit, $V_0^{}$ equals $a^3$.
  \item For the tetragonal $\upbeta$ phase, the lattice constants of our $20$-atomic model are $(\sqrt[]{2}a^{\ast},\sqrt[]{2}a^{\ast},2c^{\ast})$, thus the effective lattice constant for the single cell are $a^{\ast},c^{\ast}$, and $V_0^{}=(a^{\ast})^2c$.
  \item For the orthorhombic $\upgamma$ phase, if we denote the lattice constants of our $20$-atomic model by $(\tilde{a},\tilde{b},\tilde{c})$, where (a) the in-phase tilt is around $\tilde{\bm{c}}$, and (b) $\tilde{a}<\tilde{b}$, the effective lattice constant for a single cell are defined by $a^{\ast}=a/\sqrt[]{2},b^{\ast}=b/\sqrt[]{2},c^{\ast}=c/2$, and $V_0^{}=a^{\ast}b^{\ast}c^{\ast}$.
\end{itemize}

\begin{figure}
\includegraphics[clip,trim=1.8in 2.8in 2.4in 0.7in]{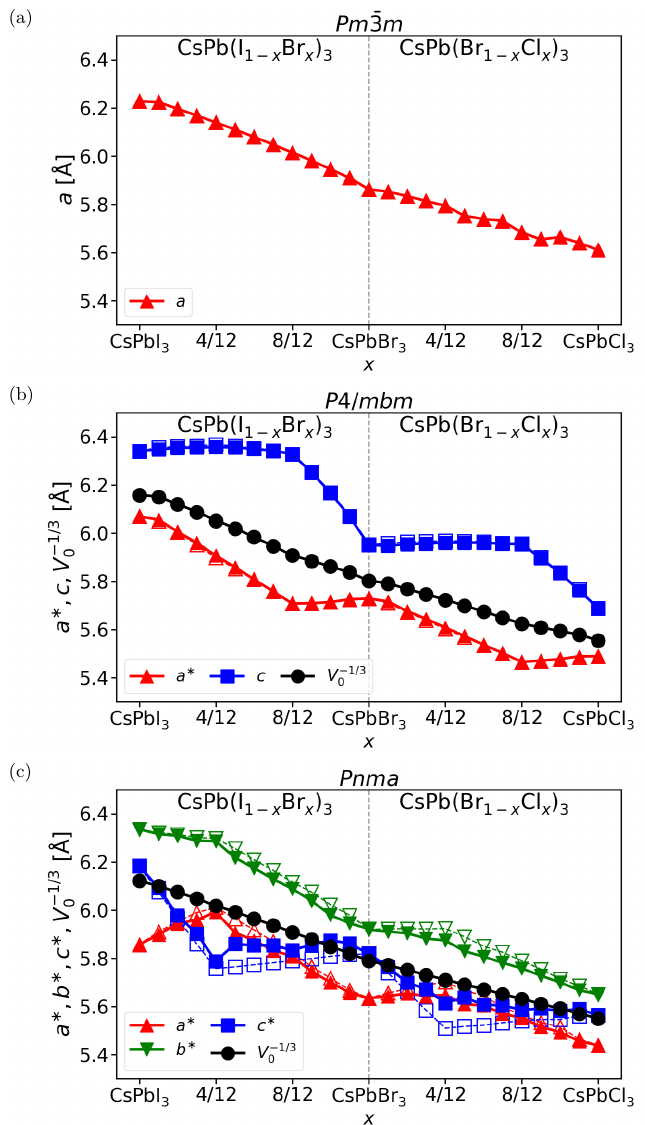}
\caption{Effective lattice constants of $\text{CsPb}(\text{I}_{1-x}^{}\text{Br}_x^{})_3^{}$ and $\text{CsPb}(\text{Br}_{1-x}^{}\text{Cl}_x^{})_3^{}$ (the left and right panel of each plot, separated by the black dashed line marking $\text{CsPbBr}_3^{}$). Shown are (a) the cubic ($\upalpha$) phase: the lattice constant $a$ (red triangles); (b) the tetragonal ($\upbeta$) phase: the effective in-plane $a^{\ast}$ (red triangles) and out-of-plane $c$ (blue squares); and (c) the orthorhombic ($\upgamma$) phase: the effective in-plane $a^{\ast}$, $b^{\ast}$, and the effective out-of-plane $c^{\ast}$ (red triangles, green triangles, and blue squares, respectively). Also shown are the cubic root of single-cell volume, $V_0^{-1/3}$ (black dots) in both (b) and (c). Data at $300$ and $0~\text{K}$ are shown by thick lines with solid marks and thin lines with empty marks, respectively.}
\label{vegard}
\end{figure}

Figure~\ref{vegard} shows how the effective lattice constants evolve with the composition of both alloy series. $\text{CsPb}(\text{I}_{1-x}^{}\text{Br}_x^{})_3^{}$ and $\text{CsPb}(\text{Br}_{1-x}^{}\text{Cl}_x^{})_3^{}$ exhibit generally similar features. For both $\upalpha$ and $\upbeta$ [Fig.~\ref{vegard}(a) and (b), respectively] phases, the results at $0~\text{K}$ and $300~\text{K}$ are similar. This finding is in accordance with that the difference between $U^{\prime}$ and $U$ (the internal energy evaluated at $0$ and $300~\text{K}$, respectively) is small [cf. Fig.~\ref{300k}(a) and (b)]. This is not surprise for the $\upalpha$ phase due to the narrow energy distribution. While for the $\upbeta$ phase, only a few low-energy configurations noticeably contribute to the ensemble at $300~\text{K}$, and their lattice constants are very similar. Differently, the difference between the $0~\text{K}$ and $300~\text{K}$ results are obvious for the $\upgamma$ phase. This means that the lattice constants of the structures that are important at $300~\text{K}$ are quite different to each other.

Vegard's law provides an important and simple method to estimate the lattice constants of alloys by linear interpolation. 
Figure~\ref{vegard} indicates that Vegard's law is appropriate only for the $\upalpha$ phase. For both non-cubic phases, only the ``quasi-cubic'' effective lattice constant $V_0^{-1/3}$ follows Vegard's law, while the individual lattice constants clearly deviates from its prediction. For the $\upbeta$ phase, upon the increase of $\text{Br}$ or $\text{Cl}$ amount in $\text{CsPb}(\text{I}_{1-x}^{}\text{Br}_x^{})_3^{}$ or $\text{CsPb}(\text{Br}_{1-x}^{}\text{Cl}_x^{})_3^{}\,$, we first observe the decrease of in-plane area, while the out-of-plane lattice constant almost remains unchanged. Beyond $x=\frac{8}{12}$, an exchange of their behavior is observed: $a^{\ast}$ rapidly decreases, while $c$ remains stable. The evolution of the $\upgamma$ phase lattice constants are more complicated, where a turning point at $x=\frac{4}{12}$ can be seen. Interestingly, both turning points are at the $x$ values where the lowest formation energies of the corresponding phases are found [cf. Fig.~\ref{energies}(b) and (c)].

\begin{figure}
\includegraphics[scale=0.32]{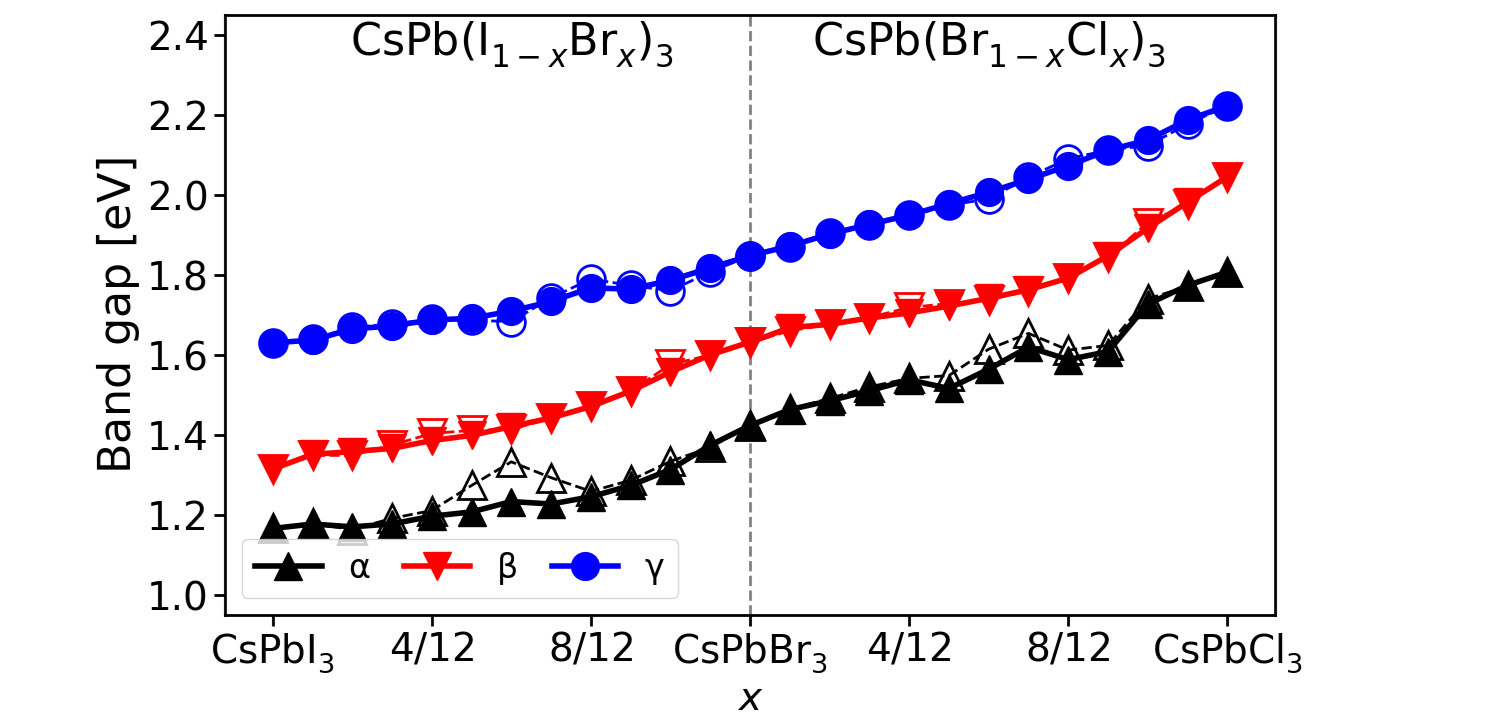}
\caption{Band gaps of $\text{CsPb}(\text{I}_{1-x}^{}\text{Br}_x^{})_3^{}$ and $\text{CsPb}(\text{Br}_{1-x}^{}\text{Cl}_x^{})_3^{}$ (the left and right panel of each plot, separated by the black dashed line marking $\text{CsPbBr}_3^{}$). Shown are cubic ($\upalpha$) phase (black triangles), tetragonal ($\upbeta$) phase (red triangles), and orthorhombic ($\upgamma$) phase (blue dots). Data at $300$ and $0~\text{K}$ are shown by thick lines with solid marks and thin lines with empty marks, respectively.}
\label{gap}
\end{figure}

Figure~\ref{gap} shows the evolution of the average band gap as a function of composition of each alloy and each phase. (Note: we only calculated band gaps with PBEsol but neither higher level theories such as hybrid functional nor spin-orbit coupling because of the computational expense. PBEsol results suffice for a qualitative analysis of trends.) As expected, the band gap continuously increases with $x$ in all three perovskite phases, confirming the possibility of tuning band gap by varying the halide-mixing amount. Interestingly, both tilted phases ($\upbeta$ and $\upgamma$) exhibit only small difference between the $0~\text{K}$ and $300~\text{K}$ results. While for the $\upalpha$ phase, large deviation is observed at $x$ around $\frac{1}{2}$, especially for $\text{CsPb}(\text{I}_{1-x}^{}\text{Br}_x^{})_3^{}\,$. This can be rationalized by that the major part of band gap contribution is narrow for both $\upbeta$ and $\upgamma$ (where the formation energy distribution is wide) phases, while it is wide for $\upalpha$ (with narrow formation energy distribution) when the amounts of $\text{I}$ and $\text{Br}$ are nearly equal.

\subsection{Phase diagram including the correction with the phonon vibrations}

\begin{figure}
\includegraphics[scale=0.32]{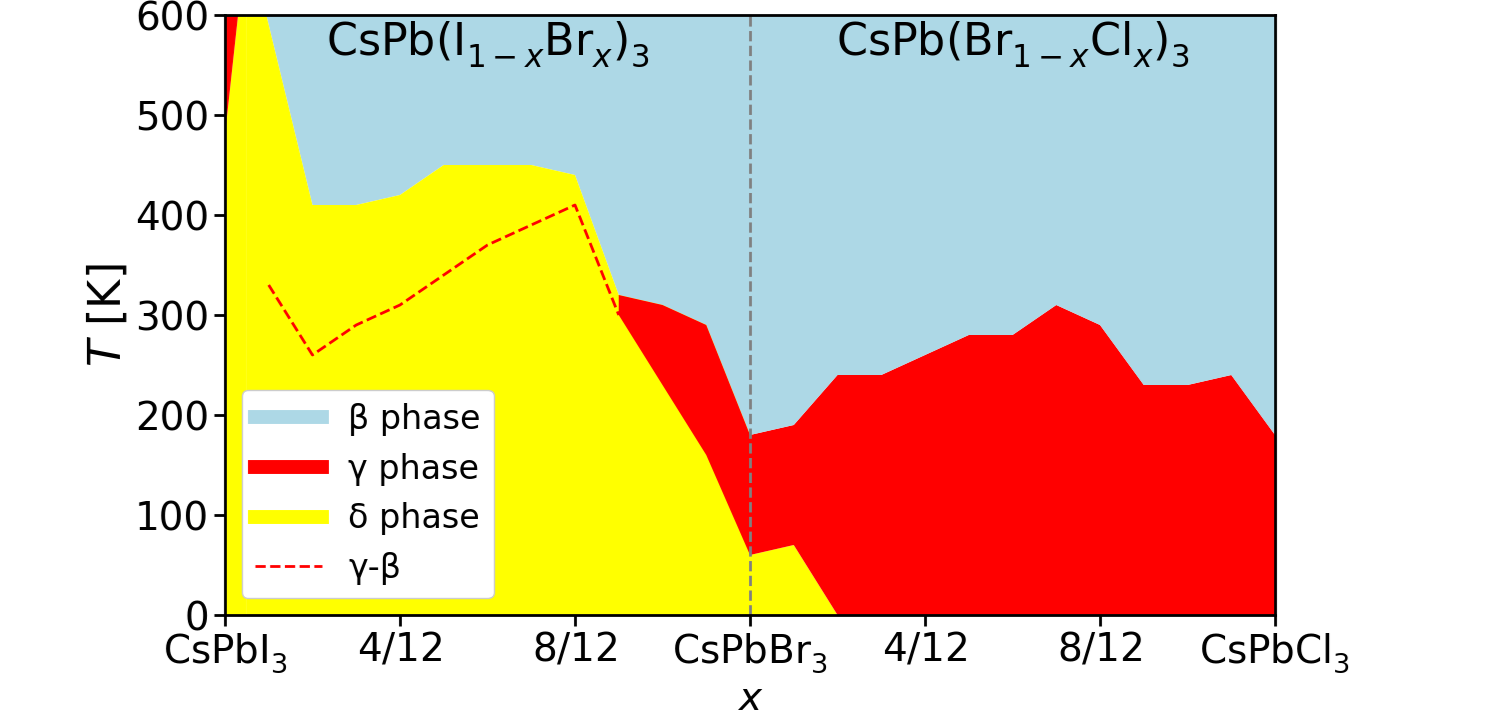}
\caption{Calculated phase diagram of $\text{CsPb}(\text{I}_{1-x}^{}\text{Br}_x^{})_3^{}$ and $\text{CsPb}(\text{Br}_{1-x}^{}\text{Cl}_x^{})_3^{}$ (the left and right panel of each plot, separated by the black dashed line marking $\text{CsPbBr}_3^{}$). The tetragonal ($\upbeta$), orthorhombic ($\upgamma$), and non-perovskite ($\updelta$) phases are colored in light blue, red, and yellow, respectively. The red dashed line indicate the $\upgamma$\--$\upbeta$ transition of $\text{CsPb}(\text{I}_{1-x}^{}\text{Br}_x^{})_3^{}$ which the composition and temperature range where the $\updelta$ phase is the most stable.}
\label{pd}
\end{figure}

With the $F(x,T)$ data of all considered phases, we can construct the temperature-composition phase diagrams of these two alloy series. To this end, we have included the contribution of phonon vibrations into our database, which effectively add a temperature-dependent term to the DFT total energy $E_i^{(0)}$ as indicated by Eqs.~(\ref{partition_total}\--\ref{single_phonon_entropy}). The results are shown in Fig.~\ref{pd}.

$\text{CsPb}(\text{Br}_{1-x}^{}\text{Cl}_x^{})_3^{}$ exhibits generally simpler phase diagram character: $\updelta\to\upgamma\to\upbeta$ as temperature increases. The non-perovskite $\updelta$ phase can only be the most stable within the low-temperature and $\text{Br}$-rich range, as can be expected by the thermodynamic state function data in Fig.~\ref{300k}. The $\upgamma$\--$\upbeta$ transition temperature ranges between $180$ and $310~\text{K}$. The calculated values are $180~\text{K}$ for both pure $\text{CsPbBr}_3^{}$ and $\text{CsPbCl}_3^{}\,$, obviously lower than the experimentally reported values ($361$ and $310~\text{K}$, respectively).

In contrast, $\text{CsPb}(\text{I}_{1-x}^{}\text{Br}_x^{})_3^{}$ exhibits much more complicated phase diagram character. For the unalloyed $\text{CsPbI}_3^{}\,$, the calculated $\updelta$\--$\upgamma$ phase transition temperature is $480~\text{K}$, considerably close to the experimental value $448~\text{K}$. $\upgamma$\--$\upbeta$ is not found in our investigated temperature range. Interestingly, the regular $\updelta$\--$\upgamma$\--$\upbeta$ profile can only be found at the $\text{Br}$-rich end. While at the $\text{I}$-rich end, once alloyed with $\text{Br}$ even with a very small amount, $\updelta$\--$\upgamma$ phase transition is not found. Instead, a direct $\updelta$\--$\upbeta$ transition is observed, whose temperature generally decreases when the amount of $\text{Br}$ increases. Our calculation results thus indicate that the $\upgamma$ ($Pnma$) phase, which is associated with the lowest total energy within the three perovskite phases for pure compounds $\text{CsPbX}_3^{}$ $\text{X}=\text{I},\text{Br},\text{I}$), is not the most stable polymorph of $\text{CsPb}(\text{I}_{1-x}^{}\text{Br}_x^{})_3^{}$ at any temperature for $x\in\big[\frac{1}{12},\frac{9}{12}\big]$. Nevertheless, we still calculated the $\upgamma$\--$\upbeta$ transition temperature within this range, which is around room temperature where the non-perovskite $\updelta$ phase is the most stable.

The phase diagram (Fig.~\ref{pd}) together with the free-energy convex hulls (Fig.~\ref{300k}) can rationalize some important experimental findings. At room temperature, the most stable phase of $\text{CsPb}(\text{Br}_{1-x}^{}\text{Cl}_x^{})_3^{}$ is either $\upgamma$ or $\upbeta$, depending on $x$. Experimentally, this perovskite alloy series can be synthesized and is generally stable within the full $x$ range \cite{Liashenko2019}. Phase separation might occur if the $\upbeta$ phase is more stable, as the convex hull of $\upbeta$ is shallow and not smooth (while it is deep and smooth for $\upgamma$). Instability is generally observed for the $\text{I}$-rich end of $\text{CsPb}(\text{I}_{1-x}^{}\text{Br}_x^{})_3^{}\,$ \cite{Beal2016,WangX2019}. For these materials, our phase diagram (Fig.~\ref{pd}) indicates that the perovskite phase is not stable at room temperature but a transition to the non-perovskite phase is preferred. This transition takes time, within it the perovskite phases (mostly $\upgamma$) experience phase separation due to the unsmooth convex hulls.

It is noteworthy that the cubic ($\alpha$) phase is not the most stable at any investigated temperature. This is due to its very high formation energy at any composition (especially considering the concave character in both alloy series), which can hardly be offset by the configuration and phonon entropies. Actually, the nature of the cubic phase of $\text{Cs}$-based perovskites are unclear. Previous computational studies indicate that it is a dynamical mixture of highly disordered tilted structures \cite{WangX2021,LiJ2023a} but not the untilted ones as used in this work. In addition, we admit that our phase diagram can only give quantitative information, mostly because the phonon calculations with harmonic approximation is not precise enough. However, including the anharmonicity effects in the phonon calculation will increase the computational demand by at least one order of magnitude, which is infeasible for the many structures considered in this paper. A much more rapid phonon calculation method would be needed, such as based on machine learning \cite{Laakso2022}.

\section{Conclusion}\label{sec:conclusion}

In summary, we used DFT to explore the whole materials space of all-inorganic $\text{CsPb}(\text{I}_{1-x}^{}\text{Br}_x^{})_3^{}$ and $\text{CsPb}(\text{Br}_{1-x}^{}\text{Cl}_x^{})_3^{}$ perovskite alloys. Thermodynamic state functions were computed based on the full dataset of alloy formation energies at different phases, compositions, and disordered configurations. Our analysis shows that configuration entropy does play an important role in stabilizing the mixed-halide perovskite alloys. Nevertheless, these effects are obviously smaller than predicted by the conventional ideal solution model in most cases, because the energies of different configurations of the same phase and composition are distributed within a large range.

This work produces a comprehensive computational dataset and thus provides a guidance of how materials stability and properties vary with composition at finite temperatures. It highlights the relevance of theoretical calculations and materials design. Most importantly, it demonstrates the necessity of having the full information of the materials space, even the high-energy configurations, to the computational estimation of properties of multi-component systems such as alloys with disordered structures.

\begin{acknowledgement}
The authors thank Patrick Rinke, Jarno Laakso, and Pascal Henkel for fruitful discussions. We acknowledge the computing resources by Xi'an Jiaotong University's HPC platform and the Computing Center in Xi'an. This work was supported by the Natural Science Foundation of Shaanxi Province of China (Grant No. 2023-YBGY-447) and National Natural Science Foundation of China (Grant No. 62281330043).
\end{acknowledgement}


\providecommand{\latin}[1]{#1}
\makeatletter
\providecommand{\doi}
  {\begingroup\let\do\@makeother\dospecials
  \catcode`\{=1 \catcode`\}=2 \doi@aux}
\providecommand{\doi@aux}[1]{\endgroup\texttt{#1}}
\makeatother
\providecommand*\mcitethebibliography{\thebibliography}
\csname @ifundefined\endcsname{endmcitethebibliography}
  {\let\endmcitethebibliography\endthebibliography}{}

\clearpage

\includegraphics[trim=1in 1in 1in 1in,clip,page= 1]{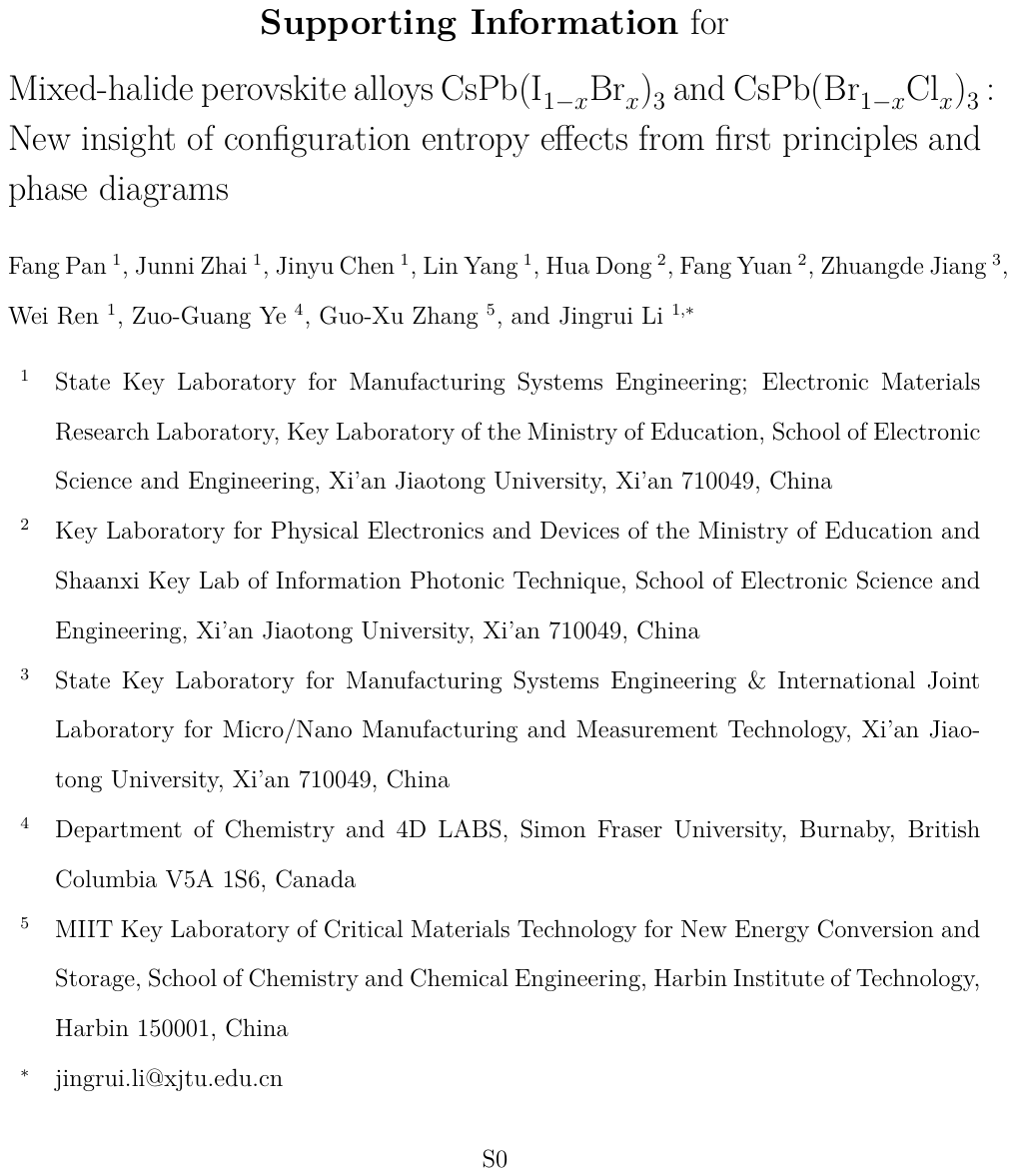}

\includegraphics[trim=1in 1in 1in 1in,clip,page= 2]{./SupportingInformation_c.pdf}

\includegraphics[trim=1in 1in 1in 1in,clip,page= 3]{./SupportingInformation_c.pdf}

\includegraphics[trim=1in 1in 1in 1in,clip,page= 4]{./SupportingInformation_c.pdf}

\includegraphics[trim=1in 1in 1in 1in,clip,page= 5]{./SupportingInformation_c.pdf}

\includegraphics[trim=1in 1in 1in 1in,clip,page= 6]{./SupportingInformation_c.pdf}

\includegraphics[trim=1in 1in 1in 1in,clip,page= 7]{./SupportingInformation_c.pdf}

\includegraphics[trim=1in 1in 1in 1in,clip,page= 8]{./SupportingInformation_c.pdf}

\includegraphics[trim=1in 1in 1in 1in,clip,page= 9]{./SupportingInformation_c.pdf}

\includegraphics[trim=1in 1in 1in 1in,clip,page=10]{./SupportingInformation_c.pdf}

\includegraphics[trim=1in 1in 1in 1in,clip,page=11]{./SupportingInformation_c.pdf}

\includegraphics[trim=1in 1in 1in 1in,clip,page=12]{./SupportingInformation_c.pdf}

\includegraphics[trim=1in 1in 1in 1in,clip,page=13]{./SupportingInformation_c.pdf}

\includegraphics[trim=1in 1in 1in 1in,clip,page=14]{./SupportingInformation_c.pdf}

\includegraphics[trim=1in 1in 1in 1in,clip,page=15]{./SupportingInformation_c.pdf}

\includegraphics[trim=1in 1in 1in 1in,clip,page=16]{./SupportingInformation_c.pdf}

\includegraphics[trim=1in 1in 1in 1in,clip,page=17]{./SupportingInformation_c.pdf}

\includegraphics[trim=1in 1in 1in 1in,clip,page=18]{./SupportingInformation_c.pdf}

\includegraphics[trim=1in 1in 1in 1in,clip,page=19]{./SupportingInformation_c.pdf}

\includegraphics[trim=1in 1in 1in 1in,clip,page=20]{./SupportingInformation_c.pdf}

\end{document}